# Chaotic and non-chaotic phases in experimental responses of a single neuron

Hagar Marmari,[1*] Roni Vardi[1*+] and Ido Kanter[1+]

[1]Gonda Interdisciplinary Brain Research Center and the Goodman Faculty of Life Sciences,
Bar-Ilan University, Ramat-Gan 52900, Israel

Consistency and predictability of brain functionalities depend on reproducible activity of a single neuron. We identify a reproducible non-chaotic neuronal phase where deviations between concave response latency profiles of a single neuron do not increase with the number of stimulations. A chaotic neuronal phase emerges at a transition to convex latency profiles which diverge exponentially, indicating irreproducible response timings. Our findings are supported by a quantitative mathematical framework and found robust to periodic and random stimulation patterns. In addition, these results put a bound on the neuronal temporal resolution which can be enhanced below a millisecond using neuronal chains.

**Introduction.-** Neuronal chaotic dynamics were exhaustively examined on a network level [1-3], mainly using simulations [4-7], but never seen experimentally in the single neuron. The possible emergence of chaotic dynamics in a single neuron is a fundamental issue since it limits the reproducibility of neuronal responses, which is essential for achieving a desired level of predictability in human brain activity [8-10]. Thus, the quantitative examination of the intrinsic chaotic behavior of a single neuron, separated from its functional neural network, is required.

Three scenarios can be theoretically expected where reproducibility is quantitatively measured by the neuronal response timings for repeated identical sets of stimulations. First, unlimited reproducibility originated from neuronal *deterministic* responses is ideal, but unrealistic, due to noisy biological environments [11, 12]. Second, neuronal response timings originate from an internal *stochastic* process [13], characterized by a small standard deviation around a biased value. Consequently, repeated stimulations of the neuron result in an additive noise which is expected to increase with square-root of the stimulation number. This minimal broadening source seems unavoidable and limits the reproducibility of neuronal behavior. Last, very poor reproducibility originates from chaotic dynamics governing responses of a single neuron. The difference in the neuronal response timings for repeated identical sets of stimulations is expected in such a chaotic dynamics to diverge exponentially with stimulation number.

In this Letter we examined the neuronal response latency, the time-gap between stimulation and evoked spike, of a neuron embedded within a large-scale network of cortical cells in vitro, but functionally separated from the network by synaptic blockers. The neuronal response latency is typically in the order of several milliseconds, and over few hundreds of periodic stimulations it shows a gradual increase which typically exceeds a millisecond [14, 15] (fig. 1(a)). For each time step, the neuronal response latency is governed by a stochastic process characterized by an increase or decrease of tens of microseconds, μs, per stimulation (fig. 1(a), inset). The probability histogram of these local changes displays a small positive bias (fig. 1(b)), which over the course of stimulation leads to the overall accumulated increase of the response latency. The average and standard deviation of this histogram quantitatively change when different portions of the latency profile are taken into account, however qualitatively they remain in the same order as in fig. 1(b).

**Non-chaotic phase.-** The reproducibility of the neuronal responses can be quantified using local and global variations between its latency profiles under *the same set of stimulation timings* (fig. 1(c)). These variations are nearly constant over hundreds of periodic stimulations and are comparable with the standard deviation ($\sigma$) of local latency changes, e.g. for $\sigma=17$ μs (fig. 1(b)) the variations are less than $3\sigma$ over 1800 stimulations (fig. 1(c)). This level of reproducibility is in contrast with a simple stochastic process, where local latency changes are independently sampled from the probability histogram in fig. 1(b). For such a stochastic process, $\sigma$ among latency profiles scales as square-root of the stimulation number, seen here to accumulate to ~0.7 ms, in contrast with the experimentally recorded value of ~0.05 ms (fig. 1(d)).

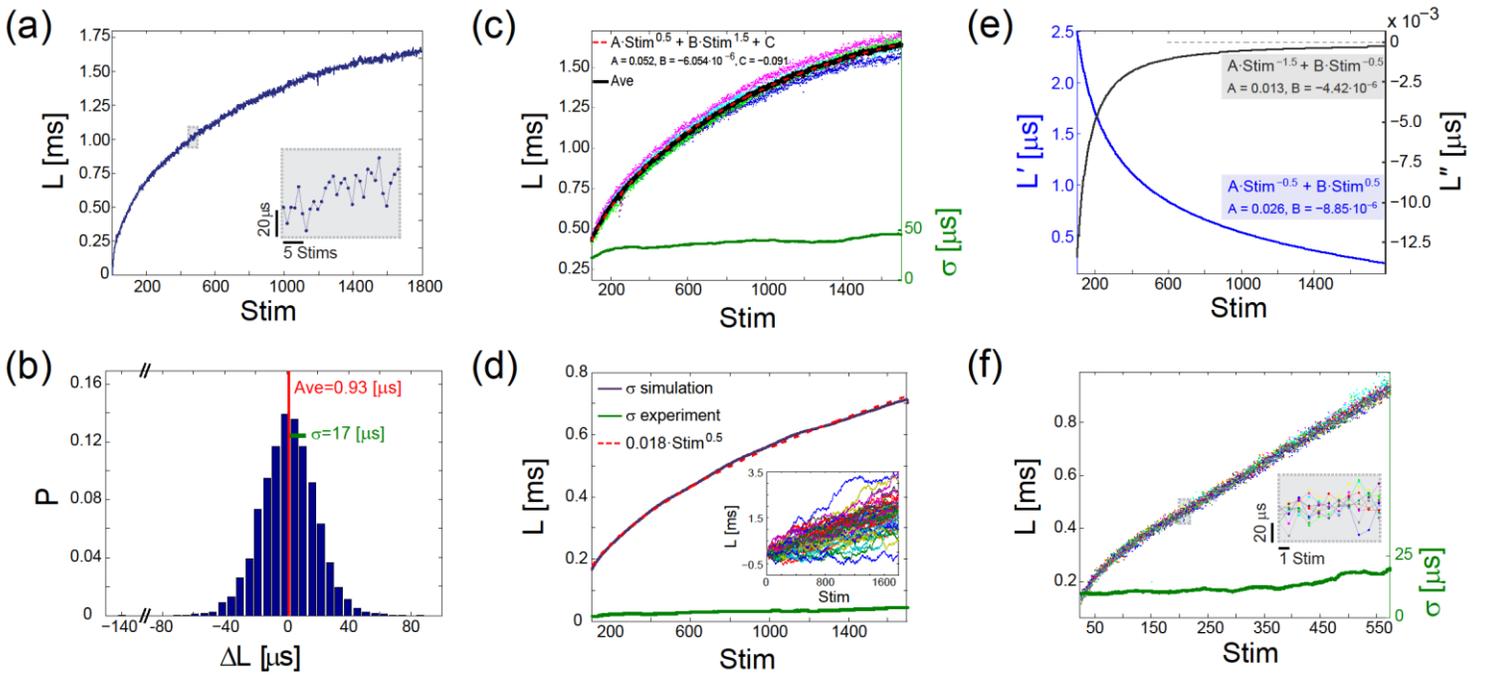

Fig. 1: (Color online). (a) The response latency of a single neuron stimulated at 10 Hz. The zoom-in (gray area) shows local changes in the neuronal response latency. Stim stands for stimulation number for this and all following figures. (b) A histogram of the difference between consecutive neuronal response latencies ΔL over 5 trials of (a), the average ΔL and standard deviation (σ) (see Methods section). (c) Neuronal response latencies of the 5 trials, their average (black) and smoothed σ (green) using 200 Stim sliding window. The approximated fit (dashed red line) after rescaling, $Stim_1=Stim/100$, results in $L=0.52 \cdot (Stim_1)^{0.5} - 6.054 \cdot 10^{-3} \cdot (Stim_1)^{1.5} - 0.091 \sim 0.52 \cdot (Stim_1)^{0.5}$ ms. (d) Simulation results of 1000 neurons (50 are exemplified, inset) whose response latency per stimulation was sampled using the histogram (b). σ of the response latencies of the simulated neurons (purple) compared to the experimental σ (green) seen in (c), both smoothed using 200 Stim sliding window. The standard deviation of the simulated trials is well fitted to $L=\sigma \cdot (Stim)^{0.5}$ ms, where $\sigma \sim 0.018$ ms. (e) L' (blue) and L'' (black) computed from the fit seen in (c). (f) Response latencies of a single neuron stimulated at random time-lags in the range of [95, 105] ms, over 10 trials, and their smoothed σ (green) using 50 Stim sliding window. A zoom-in (gray area) exemplifies the local variability.

This supreme reproducibility is a result of the concave average latency profile, i.e. $L'' \equiv d^2L/dStim^2 < 0$ (fig. 1(c) and 1(e)), over several trials with similar stimulation profiles. For illustration, let us compare the time evolution of two initially close neuronal response latencies, $L_1$ and $L_2=L_1+\gamma$, $(L_2-L_1=\gamma)$. Following a stimulation, the latencies change to

$$L_1+L'(L_1)$$

and

$$L_2 = L_1+\gamma+L'(L_2)$$

in accordance with the first derivative of the latency profile (fig. 1(e), $L' \equiv dL/dStim$). Now the difference between the two nearby latencies is

$$\gamma+L'(L_2)-L'(L_1)$$

Since the latency profile is concave, $L'(L_2)-L'(L_1)<0$, the difference between two nearby trajectories around the average latency decreases and can be expressed as

$$\gamma + L'(L_1+\gamma) - L'(L_1) = \gamma + \gamma L''(L_1) = \gamma(1 + L''(L_1))$$

where the negative constant L'' represents an effective intermediate second derivative in this concave region. Now, an iterative process leads to a multiplicative effect

$$\Delta_{Stim} = \gamma(1 + L'')^{Stim} = \gamma e^{\ln(1 + L'')Stim}$$

where $\Delta_0=\gamma$. The negative Lyapunov exponent, $\ln(1+L'')<0$, indicates an exponential convergence of two nearby trajectories, allowing to overcome the inherent broadening of a stochastic process. This non-chaotic phase, represented by a concave neuronal response latency profile, generally found in our experiments to scale in the leading order as $Stim^{0.5}$ (fig. 1(c)).

This process is very similar to a Bernoulli map [16, 17], $X_{n+1}=(aX_n)$ mod 1, which is non-chaotic for $a<1$.

The variable $X_n$ stands not for the neuronal response latency L, but for L'($L_n$). The simplification of the Bernoulli map is that $a$-1 (i.e. L'') is a constant, independent of $L_n$, whereas L'' for the neuronal response latency varies as a function of $L_n$.

The supreme reproducibility seen under periodic stimulations (fig. 1(c)) remains robust even under random stimulation patterns (fig. 1(f)). The same neuron, receiving different stimulation patterns sampled at random from a given distribution, shows, as expected, local variations between its latency profiles (fig. 1(f), inset). However, on the global scale these variations produce a nearly constant σ over many hundreds of stimulations.

**Chaotic phase.-** The concave average latency profile is typically followed by a convex average latency profile preceding the intermittent period [14, 18] (fig. 2(a)), a transition which substantially varies among neurons. Qualitatively, this transition is accompanied by a rapid increase in σ among slightly perturbed latency profiles around the average profile (fig. 2(a)), seemingly stemming from the point of transition; however, a quantitative analysis is still demanded. Initially, and far from the transition to the convex latency profile, the neuronal response latency displays a concave profile which scales as $Stim^{0.5}$ (fig. 2(b)), similarly to fig. 1(c). The entire averaged increase of the neuronal response latency, excluding the initial concave-like profile, was found to be well approximated by a cubic polynomial (fig. 2(b)). The derivatives for this scaled fit quantitatively pinpoint the critical stimulation, $Stim_C$, at which the transition from concave to convex latency profiles occurs (fig. 2(c)). Since in the convex region ($Stim>Stim_C$), L''>0, the Lyapunov exponent is positive, ln(1+L'')>0, and an exponential divergence between nearby latency profiles is expected. Experimentally, the latency difference between two nearby trajectories fits better to an exponential divergence (linear fit between ln($\Delta_{Stim}$)~ln(σ) and Stim-$Stim_C$) than to a power-law (fig. 2(d), inset), demonstrating that this is a chaotic process. Both the theoretical arguments and the experimental data indicate that the convex profile represents a new chaotic neuronal phase.

The number of stimulations until the emergence of the chaotic phase depends on the stimulation rate (fig. 3(a)). Nevertheless, a support for a *universal behavior* is found where the transition to a chaotic phase is roughly a function of the latency increase, independent of the stimulation rate (fig. 3(b), blue line).

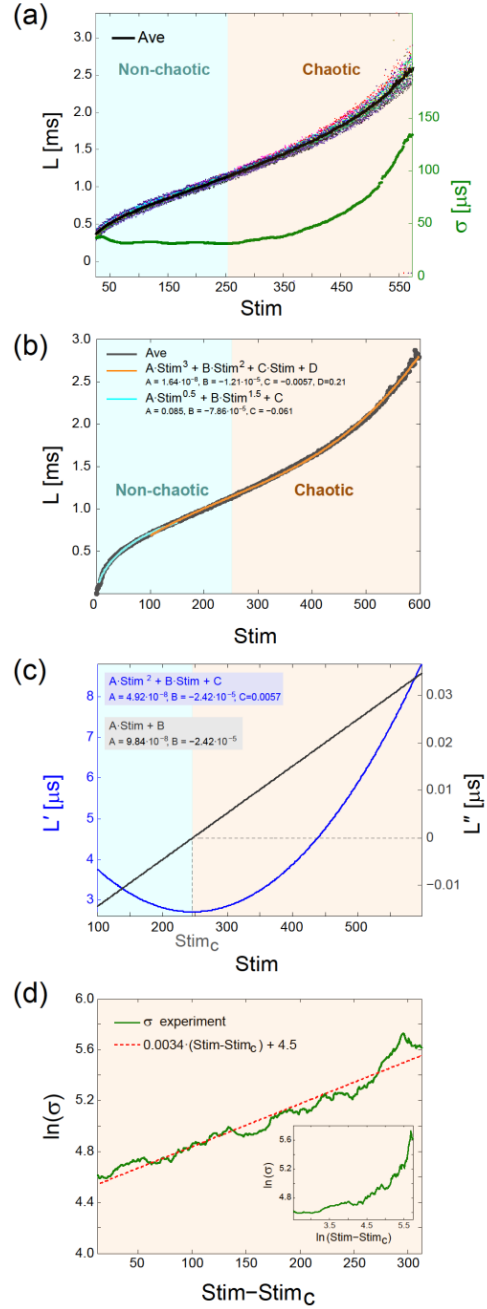

Fig. 2: (a) Response latencies of a single neuron, stimulated at 20 Hz over 15 trials, their average (black) and smoothed σ (green) using 50 Stim sliding window. (b) The average of the neuronal response latencies seen in (a) (gray). The approximated fit for Stim∈[50,150] (teal) indicates, after rescaling, a dominating behavior L~0.85·(Stim/100)$^{0.5}$ ms. For Stim∈[100,600] the latency is well approximated by a cubic polynomial fit (orange). (c) L' (blue) and L'' (black) computed from the cubic polynomial fit in (b), where L''=0 at $Stim_C$=247. (d) A linear fit (dashed red-line) for ln(σ) versus Stim-$Stim_C$ (green), indicating a chaotic behavior with a Lyapunov exponent of 0.0034. The inset, ln(σ) versus ln(Stim-$Stim_C$), excludes a power-law fit.

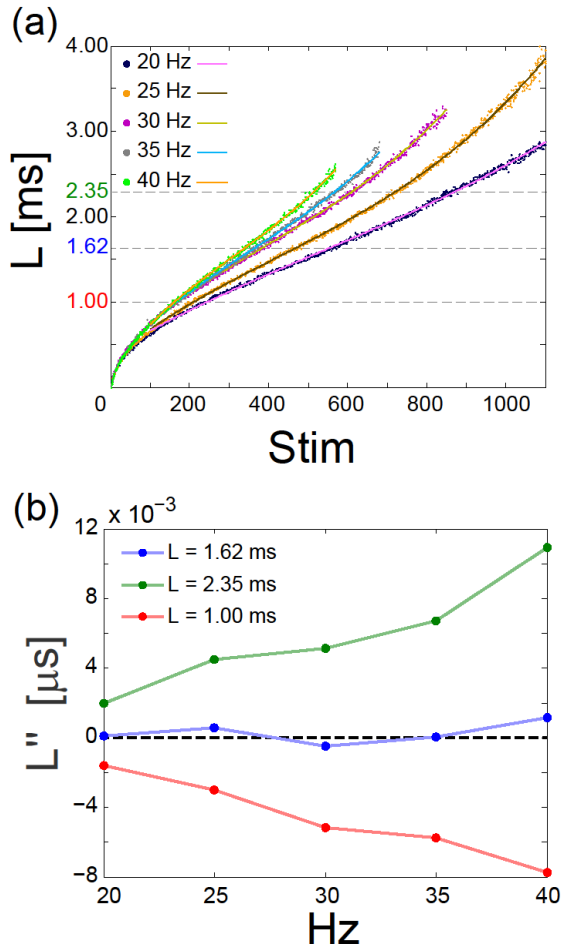

Fig. 3: (a) Response latencies of a single neuron, stimulated at various frequencies (colored dots) and their cubic polynomial fit for Stim>100 (full lines). (b) L" obtained from the fit in (a) for different stimulation frequencies at response latencies 1.00 (red), 2.35 (green) and 1.62 (blue) ms.

**Neuronal temporal resolution**.- We demonstrated here an extremely robust feature of reproducibility in the responses of a single neuron, which was experimentally verified under both periodic and random stimulation patterns. This supreme reproducibility hints at the temporal resolution of neuronal responses, where during the concave (non-chaotic) phase of the neuronal response latency, one can identify two nearby stimulation rates from the knowledge of their non-overlapping neuronal latency profiles. The neuronal temporal resolution is a much investigated topic [19-21], although fundamental questions still remain unanswered. In the auditory system, for example, a microsecond time resolution is needed while neuronal spiking resolution is in the millisecond range, raising the question whether such a highly precise neuronal temporal code is possible.

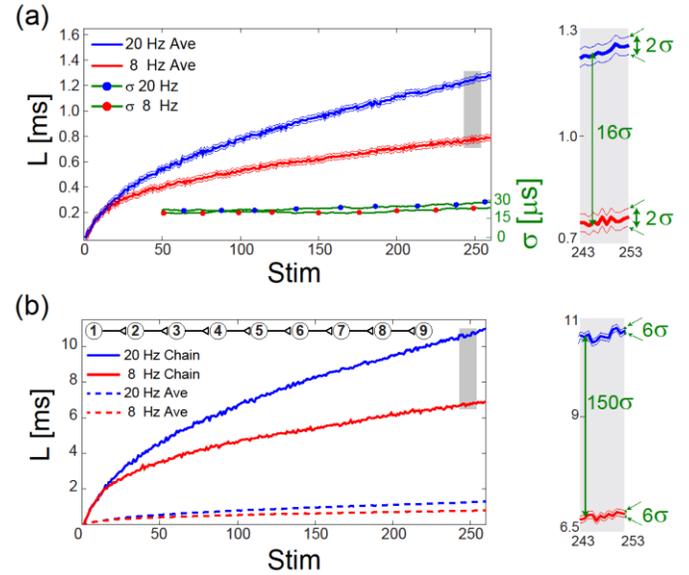

Fig. 4: (a) Average response latencies obtained from 5 trials of a single neuron stimulated at 20 Hz (blue) and 8 Hz (red), and their smoothed σ using 50 Stim sliding window. The zoom-in (gray area) shows 2σ broadening. The ratio 16σ/2σ=8 indicates that the neuronal temporal resolution between 8 Hz (125 ms) and 20 Hz (50 ms) is (125-50)/8~9 ms. (b) Schematic of a neuronal chain consisting of N=9 neurons (top), and its accumulated response latency at 20 Hz (blue) and 8 Hz (red). The average response latency for N=1 (dashed lines, from (a)) is shown for comparison. The zoom-in (gray area) shows a 150σ latency gap, whereas the estimated broadening is 2σ·sqrt(9)=6σ. Their ratio, 150σ/6σ=25, indicates (125-50)/25=3 ms chain temporal resolution.

Using the neuronal response latency, the neuronal temporal resolution can be approximated by the ratio between the latency gap between two latency profiles and σ. For illustration, we examine the ratio between the latency profiles of a neuron stimulated at 8 and 20 Hz at its non-chaotic phase, and their variability, which results in a temporal resolution of ~9 ms (fig. 4(a)). Typically, σ was found to be frequency independent and the latency gap maximized after a large number of stimulations in the non-chaotic phase (fig. 4(a)). Consequently, the temporal resolution is expected to be enhanced towards a millisecond with increase of the stimulation frequency. This experimentally calculated temporal resolution fits well with past results [19-21], but still cannot clarify the feasibility of sub-millisecond time resolution.

In order to enhance the temporal resolution, a neuronal chain [22] can be suggested (as in fig. 4(b), top). The latency gap in such a chain resulting from a pair of stimulation rates accumulates and increases linearly with the number of neurons constituting the chain, N (fig. 4(b)). The accumulated standard deviation of the chain, $\sigma \cdot \sqrt{N}$, is, however, a sum of the independent deviations of each neuron in the chain. $\sigma$ stands for the standard deviation of a single neuron (e.g. fig. 4(a)). Consequently, the ratio between these two factors in a chain results in an enhanced time resolution by a factor of $\sqrt{N}/N = 1/\sqrt{N}$ (fig. 4(b)). Here, the experimental results show that for a chain of 9 neurons, the ratio between the chain's latency profile at 8 and 20 Hz and its $\sigma$ is ~3 ms, and is expected to decrease below a millisecond for higher stimulation frequencies. Hence, longer neuronal chains can refine the neuronal temporal code to much below a millisecond, which results in a frequency resolution of hundredths of Hz.

We experimentally demonstrated the emergence of a chaotic phase in the dynamics of a single neuron, using the reproducibility of the neuronal responses to the same stimulation pattern. This phase is characterized by a single positive Lyapunov exponent, and the transition between the non-chaotic and chaotic phases may be governed by several universal features. However, future research is required in order to understand the cellular mechanisms underlying these phenomena, as well as a generalization to dynamics under cell assemblies.

**Methods.-** Culture preparation, synaptic blockers, stimulation and recording, cell selection, stimulation control, data analysis and spike detection were performed as described in previous publications [15, 22].

*Histogram of local neuronal response latency changes.* A neuron was stimulated 1800 times at 10 Hz, over 5 trials. For each trial, the difference in the neuronal response latency per step was computed ($\Delta L_i = L_i - L_{i-1}$, where $L_i$ is the latency at the *i*th stimulation. This data was used to generate a histogram of $\Delta L_i$ consisting of 40 bins.

*Neuronal response latency simulation.* 1000 neurons were simulated for 1800 steps such that each $\Delta L_i$ was sampled from the experimental $\Delta L_i$ histogram. For each simulated step, a random number was selected from a uniform distribution, $U \sim [0,1]$. The accumulated probabilities of the histogram bins were computed, and $\Delta L_i$ was chosen from the bin with the largest accumulated probability smaller than the random number. For each trial, the simulated $\Delta L_i$ were summed to show the accumulated latency. The first 50 simulated neurons are displayed in the inset of fig. 1(d).

**Acknowledgements.-** We thank Moshe Abeles for stimulating discussions. Invaluable computational assistance by Yair Sahar, Amir Goldental and technical assistance by Hana Arnon are acknowledged. This research was supported by the Ministry of Science and Technology, Israel.

* These authors equally contributed to this work
+ ido.kanter@biu.ac.il, ronivardi@gmail.com